\RequirePackage{fix-cm}

\documentclass[smallextended]{svjour3}       
\smartqed  
\usepackage{graphicx}
\usepackage{color}
\usepackage{kantlipsum, wrapfig,graphicx}
\usepackage{amsmath}
\usepackage{amssymb} 
\usepackage[numbers,sort&compress]{natbib}
\begin{document}

\title{Electroweak Currents from Chiral Effective Field Theory
\thanks{SP and GBK are supported by the U.S.~Department of Energy under contract DE-SC0021027, 
and through the Neutrino Theory Network (SP) and the FRIB Theory Alliance award DE-SC0013617 (SP). 
AB is supported by the U.S.~Department of Energy Early Career Award program, Office of Science,  Office of Nuclear Physics.
}
}


\author{Alessandro Baroni      \and
        Garrett B. King \and Saori Pastore 
}


\institute{A. Baroni \at
              Theoretical Division, Los Alamos National Laboratory, Los Alamos, NM 87545, USA \\
              \email{abaroni@lanl.gov}           
           \and
           G. B. King \at
            Department of Physics and the McDonnell Center for the Space Sciences at Washington University in St. Louis, MO, 63130, USA\\
              \email{kingg@wustl.edu}     
           \and
           S. Pastore \at
              Department of Physics and the McDonnell Center for the Space Sciences at Washington University in St. Louis, MO, 63130, USA\\
              \email{saori@wustl.edu}
}

\date{Received: date / Accepted: date}

\maketitle

\begin{abstract}
Since the pioneering work of Weinberg, Chiral Effective Field Theory ($\chi$EFT) has been  
widely and successfully utilized in nuclear physics to study many-nucleon interactions and associated electroweak currents.
Nuclear $\chi$EFT has now developed into an intense field of research and is applied to study light to medium mass nuclei. 
In this contribution, we focus on the development of electroweak currents from 
$\chi$EFT and present applications to selected nuclear electroweak observables.
\keywords{Chiral Effective Field Theory \and Nuclear Structure \and Nuclear Reactions}
\end{abstract}

A major objective of nuclear theory is to explain the structure and dynamics of nuclei and their interaction with electroweak probes
in a fully microscopic approach.  In such an approach, nucleons interact with each 
other via many-body (primarily, two-and  three-nucleon)  interactions, 
and  with  external  electroweak  probes, such as electrons, neutrinos, and photons,  
via   many-body currents  describing the couplings of these probes to individual
nucleons and many-body clusters of correlated nucleons. Over the past sixty years, several
highly accurate phenomenological interactions~\cite{Wiringa:1994wb,Machleidt:2000ge} have been developed and successfully applied to 
study nuclear properties. Despite this success, phenomenological theories are hardly improvable;
moreover, their connection to the underlying theory ultimately governing nuclear dynamics, 
that is Quantum Chromodynamics (QCD), is ambiguous. Chiral effective field  theory
($\chi$EFT), formulated by Weinberg
in the Nineties~\cite{Weinberg:1990rz,Weinberg:1991um,Weinberg:1992yk}, resolves these shortcomings. 
\begin{wrapfigure}{r}{0.5\textwidth}
  \raisebox{0pt}[\dimexpr\height-0.6\baselineskip\relax]{  \includegraphics[width=.46\textwidth]{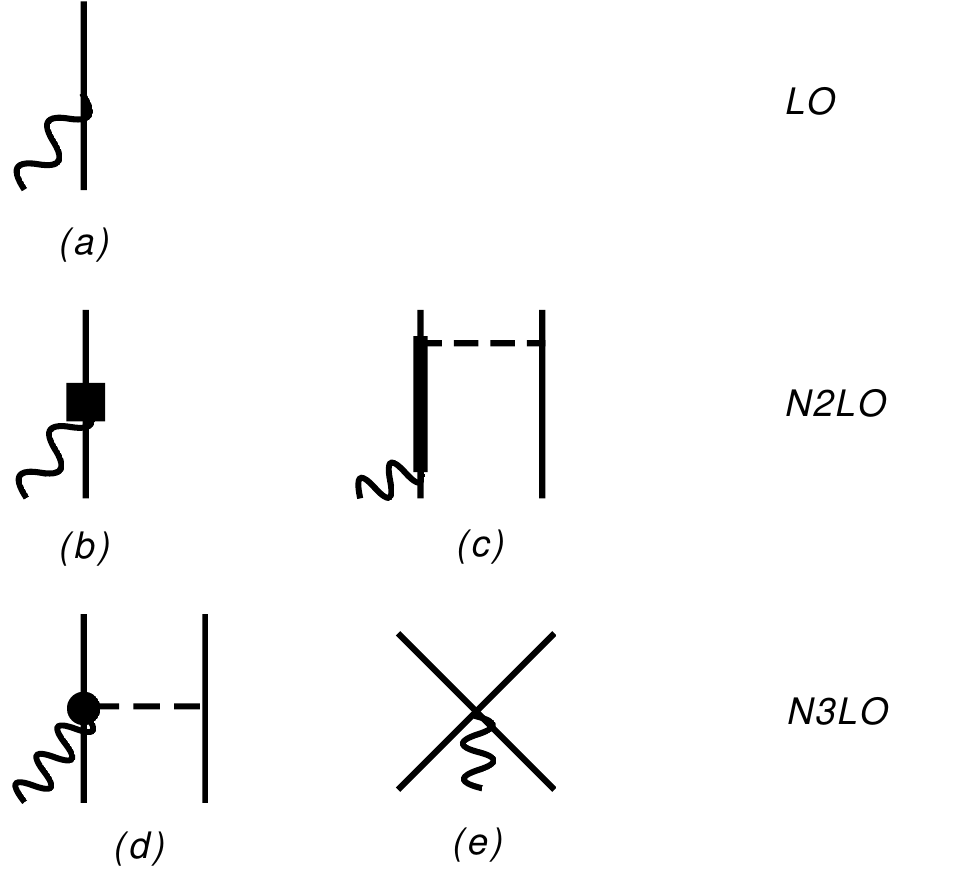}}
  \caption{Diagrams illustrating the contributions to the axial current up to 
N3LO. Nucleons, deltas, pions, and external weak fields are denoted by solid, thick-solid, 
dashed, and wavy lines, respectively. The square in panel b represents relativistic 
corrections, while the dot in panels d denotes a vertex induced by subleading terms 
in the $\pi$-nucleon chiral Lagrangian~\cite{Baroni:2016xll}.}\label{fig:dia}
\vspace{-0.3cm}
\end{wrapfigure}
$\chi$EFT is a low-energy approximation of QCD whose degrees of
freedom are bound states of QCD ({\it e.g.}, pions, protons, neutrons, etc.). It
exploits the symmetries exhibited by QCD in the low-energy regime,
in particular chiral symmetry, to constrain and determine the interactions of 
pions among themselves and with other baryons.  The pion couples to these particles by powers
of its momentum $Q$, and the Lagrangians describing these
interactions can be expanded in powers of $Q/\Lambda$, where $\Lambda\sim1$ GeV
represents the chiral-symmetry breaking scale and characterizes
the convergence of the expansion. Therefore, the validity of
the theory is confined to kinematic regions where the constraint
$Q\ll \Lambda$ is realized. The coefficients of the chiral expansion, or
low-energy constants (LECs), are unknown and need to be fixed
by comparison with experimental data or calculated by nonperturbative QCD 
computational methods such as Lattice QCD~\cite{Aoki:2016frl,Bijnens:2014,Orginos:2015aya,Shanahan:2016,Detmold:2021,Beane:2014ora,Savage:2016kon,Tiburzi:2017,Davoudi:2020,Cirigliano:2020,Parreno:2021}.

This extremely powerful approach provides an expansion of the Lagrangians in powers of a small
momentum as opposed to an expansion in the strong coupling constant,
restoring {\it de facto} the applicability of perturbative techniques also in the
low-energy regime. Due to the chiral expansion it is then  possible, in principle,
to evaluate an observable to any degree of desired accuracy and
to know {\it a priori} the hierarchy of interactions contributing to
the (low-energy) process under study. The systematic expansion in 
$Q/\Lambda$ naturally arranges the operators in increasing 
numbers of nucleons. For example, in the $Q/\Lambda$  power counting, 
three-nucleon forces are suppressed with respect to two-nucleon 
forces, and so on. Another crucial feature of $\chi$EFT is that
many-body electroweak currents can be readily and consistently constructed 
within the same $\chi$EFT adopted to derive the many-nucleon interactions. 

Since the pioneering work of Weinberg~\cite{Weinberg:1990rz,Weinberg:1991um,Weinberg:1992yk},
this calculational scheme has been widely utilized in nuclear physics~\cite{Binder:2018pgl,Epelbaum:2008ga,Machleidt:2011zz,Entem:2014msa,Entem:2003ft,Epelbaum:2002vt,Epelbaum:2014sza,Gezerlis:2014zia,Epelbaum:1998ka,Epelbaum:1999dj,Kaiser:1997mw,Krebs:2007rh,Krebs:2012yv,Krebs:2018jkc,Ordonez:1995rz,vanKolck:1994yi,Piarulli:2014bda,Piarulli:2016vel,Piarulli:2017dwd,Gandolfi:2020pbj,10.3389/fphy.2019.00245,Ekstrom:2013kea,Ekstrom:2017koy,Park:1993jf,Park:1995pn,Park:1998wq,Park:2002yp,Phillips00,Phillips2003,Phillips05,Walzl2001,Pastore:2008ui,Pastore:2009is,Pastore:2011ip,Piarulli:2012bn,Kolling:2009iq,Kolling:2011mt,Krebs:2016rqz,Krebs:2020rms,Baroni:2015uza,Baroni:2016xll,Klos:2016omi,Song:2007,Song:2009,Schiavilla:2018udt} and
nuclear $\chi$EFT has developed into an intense field of research. 
In this contribution, we focus on the development of electroweak currents from 
$\chi$EFT~\cite{Park:1995pn,Park:1998wq,Park:2002yp,Phillips00,Phillips2003,Phillips05,Walzl2001,Pastore:2008ui,Pastore:2009is,Pastore:2011ip,Piarulli:2012bn,Kolling:2009iq,Kolling:2011mt,Krebs:2016rqz,Krebs:2020rms,Baroni:2015uza,Baroni:2016xll,Klos:2016omi} and present applications to selected observables in light and medium mass nuclei~\cite{Bacca:2014tla,kolling2012,Rozpedzik,Gysbers:2019uyb,Pastore:2012rp,Pastore:2014oda,Pastore:2017uwc,King:2020wmp}.

Nuclear electroweak current ({\bf j}) and charge ($\rho$) operators can be expressed as a sum of 
one and many-body contributions, namely
\begin{eqnarray}
\rho({\bf q})&=&\sum_i \rho_i({\bf q}) +\sum_{i<j} \rho_{ij}({\bf q})+\cdots \nonumber\, , \\
{\bf j}({\bf q})&=&\sum_i {\bf j}_i({\bf q}) +\sum_{i<j} {\bf j}_{ij}({\bf q})+\cdots \nonumber\, ,
\end{eqnarray}
where {\bf q} is the momentum carried from the external field and the dots denote three-body operators and beyond.
To illustrate the kind of currents emerging from a $\chi$EFT with pions, nucleons, and
delta excitations, in Fig.~\ref{fig:dia} we show the vector axial current 
at tree-level up to next-to-next-to-next-to-leading order (N3LO) in the chiral expansion at zero momentum transfer.  
The major contribution is from the leading order single-nucleon operator (panel a), this corresponds
to the standard Gamow-Teller operator. Two-body corrections enter at N2LO with a transition current
where a nucleon is excited into a delta, which decays into a pion that gets reabsorbed by a second 
nucleon (panel c). At N3LO, there is another one-pion range two-body current (panel d), along with a contact current (panel e).  
Because of the power counting, kinematic effects can also be accounted for and arranged within the power expansion. 
A term of this kind is illustrated in panel b, representing a relativistic correction to the leading one-body operator. 
All the LECs entering these operators are experimentally known except for that entering the contact 
term in panel e, which is determined by fits to experimental data~\cite{Baroni:2018fdn}.

\begin{wrapfigure}{l}{0.5\textwidth}
  \raisebox{0pt}[\dimexpr\height-0.6\baselineskip\relax]{  \includegraphics[width=0.48\textwidth]{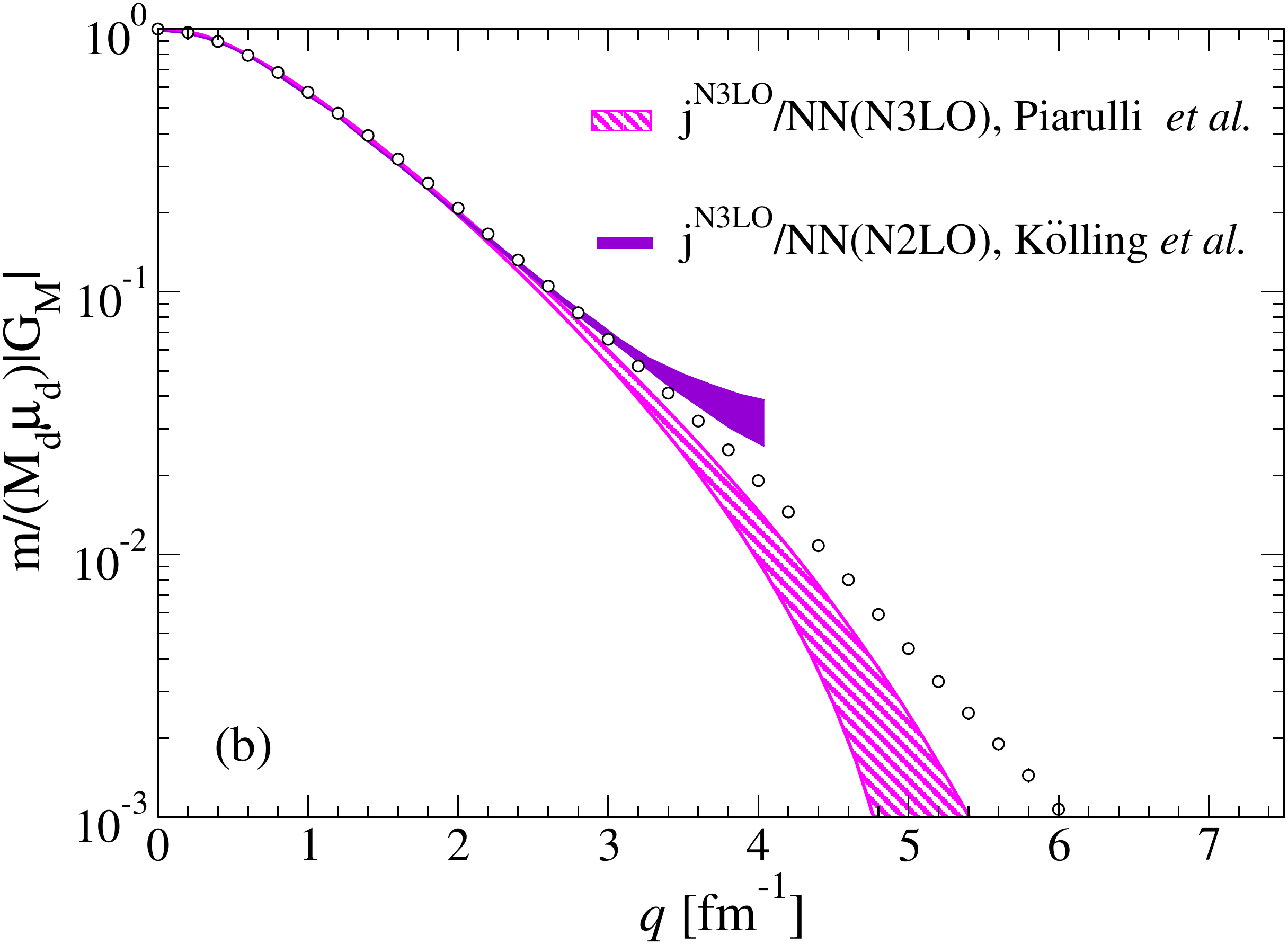}}
  \caption{Calculated deuteron magnetic form factor from Ref.~\cite{Piarulli:2012bn} (magenta
hatched band) compared with results from Ref.~\cite{kolling2012} (purple solid band). Experimental
data are from Refs.~\cite{STANFORD65,MAINZ81,SACLAY85,BONN85,SLAC90,SICK2001}}\label{fig:ff}
\vspace{-0.5cm}
\end{wrapfigure}

The pioneering derivations of both vector and axial vector currents up to one-loop contributions are from
Park, Min and Rho~\cite{Park:1993jf,Park:1995pn} who worked within covariant perturbation theory with pions and 
nucleons as degrees of freedom. These currents, have been used by Song and collaborators
in several hybrid calculations in $A\le3$ nuclei~\cite{Song:2007,Song:2009}. The electromagnetic
charge operator was first investigated within $\chi$EFT by Walzl {\it et al.} in
Ref.~\cite{Walzl2001} and Phillips in Refs.~\cite{Phillips00,Phillips2003,Phillips05}. More recently, derivations based on two different 
implementations of time-ordered perturbation theory appeared in the literature. 
One is from the so-called Pisa-JLab group~\cite{Pastore:2008ui,Pastore:2009is,Pastore:2011ip,Piarulli:2012bn} and the other is from the Bochum-Bonn group~\cite{Kolling:2009iq,Kolling:2011mt,Krebs:2019}. 
The latter is based on the method of the unitary transformation~\cite{Okubo:1954} that
decouples, in the Hilbert space of pions and nucleons,  states consisting of nucleons only from those including both pions and nucleons.
The two derivations differ in the treatment of reducible diagrams~\cite{Pastore:2009is,Kolling:2011mt,Krebs:2016rqz,Baroni:2018fdn}. When calculating 
box diagrams entering the electromagnetic charge and current operators~\cite{Pastore:2011ip,Krebs:2020rms}, 
the two methods lead to results that are in agreement.  However,  as discussed at 
length in Refs.~\cite{Baroni:2018fdn,Krebs:2020rms}, the two groups find different results for the box diagrams in 
the loop contributions to the axial current operator. The (minor) numerical impact of this 
difference has been investigated in Ref.~\cite{Baroni:2018fdn}. Further checks are underway to clarify 
the origin of these differences ~\cite{Krebs:2020box}. Both formulations, have been used in calculations of 
electroweak observables in (primarily) light nuclei.

\begin{figure}[!t]
    \centering
    \begin{minipage}{.48\textwidth}
  \raisebox{0pt}[\dimexpr\height-0.6\baselineskip\relax]{\includegraphics[width=1.0\textwidth]{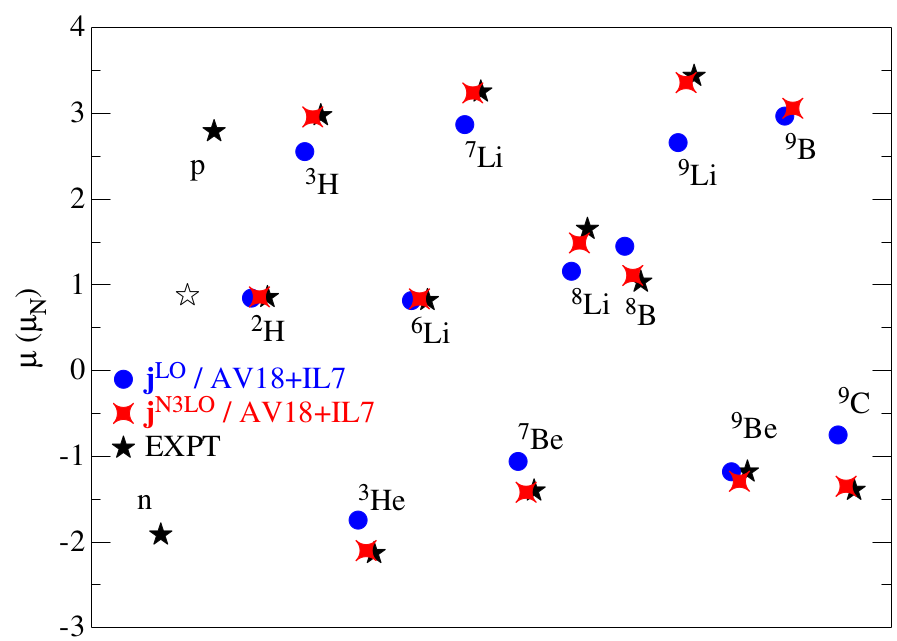}}
  \caption{Magnetic moments in nuclear magnetons for $A\le9$ nuclei from
Ref.~\cite{Pastore:2014oda}. Black stars indicate the experimental values from Refs.~\cite{Borremans05,Tilley02,Tilley04}, while
blue dots (red diamonds) represent Green's Function Monte Carlo calculations which include the LO one-body currents (one-body plus two-body currents at N3LO)
from $\chi$EFT. }\label{fig:mm}
    \end{minipage}%
    \hspace{0.1cm}
    \begin{minipage}{.48\textwidth}
     \raisebox{0pt}[\dimexpr\height-0.6\baselineskip\relax]{\includegraphics[width=1.\textwidth]{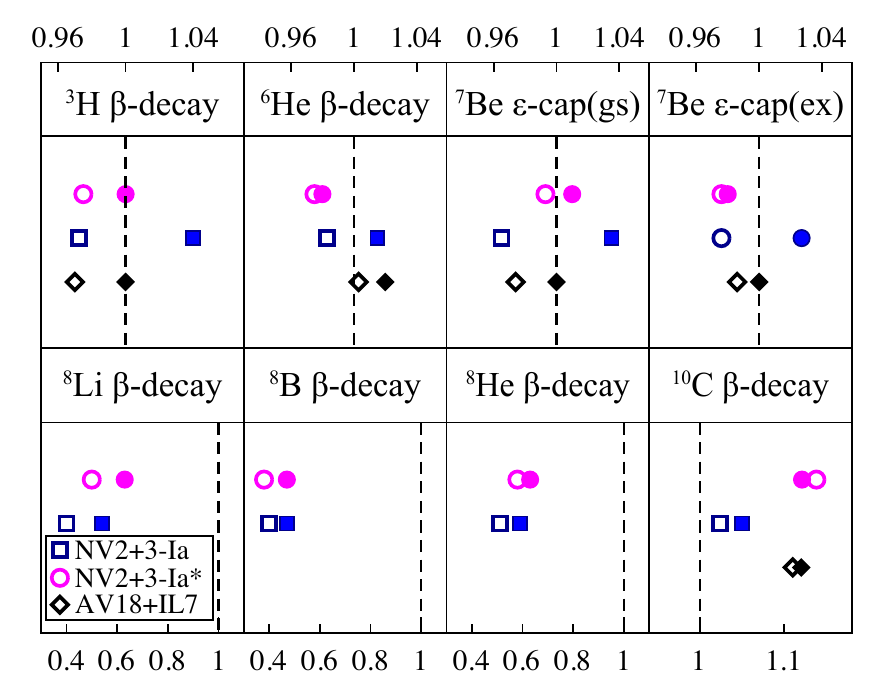}}
  \caption{Ratios of Green's function Monte Carlo calculations to experimental values of the Gamow-Teller reduced matrix elements in the $^3$H, $^6$He, $^7$Be, $^8$B, $^8$Be, $^8$He and $^{10}$C weak transitions from Refs.~\cite{Pastore:2017uwc,King:2020wmp}. Theory predictions correspond to the $\chi$EFT axial current at LO (empty symbols) and up to N3LO (filled symbols).  }\label{fig:beta}
    \end{minipage}
\end{figure}

In Fig.~\ref{fig:ff}, we show calculations of the deuteron magnetic form factor based on
$\chi$EFT currents calculated by K\"olling {\it et al.} in Ref.~\cite{Kolling:2009iq} and Piarulli {\it et al.}
in Ref.~\cite{Piarulli:2012bn}. The theoretical results are in very good agreement with each other and with
the experimental data for values of momentum transferred $q \lesssim 3$ fm$^{-1}$.
$\chi$EFT currents are first used for nuclei with $A>3$ in Ref.~\cite{Pastore:2012rp} where they are included 
when studying magnetic moments and electromagnetic transitions in $A\le10$ systems. 
Two-body currents always improve on the agreement between experimental data and
theoretical calculations. A long standing under-prediction~\cite{Utsuno:2004wn} by less sophisticated
theoretical estimations of the measured $^9$C magnetic moment is in Ref.~\cite{Pastore:2012rp}
explained by the presence of a 40\% correction generated by two-body electromagnetic currents.
This enhancement can be appreciated  in Fig.~\ref{fig:mm} by comparing blue dots (representing calculations based on the single nucleon paradigm)
and red diamonds (representing calculations with two-body electromagnetic currents).

Axial currents are tested primarily in beta decays and electron capture processes 
for which data are readily available and known for the most part with great accuracy. 
The long-standing problem of the systematic overprediction of Gamow-Teller beta decay matrix elements~\cite{Chou:1993zz}
in simplified nuclear calculations, the so-called `$g_A$ problem', has been recently
addressed by several\begin{wrapfigure}{l}{0.5\textwidth}
  \raisebox{0pt}[\dimexpr\height-0.6\baselineskip\relax]{  \includegraphics[width=0.48\textwidth]{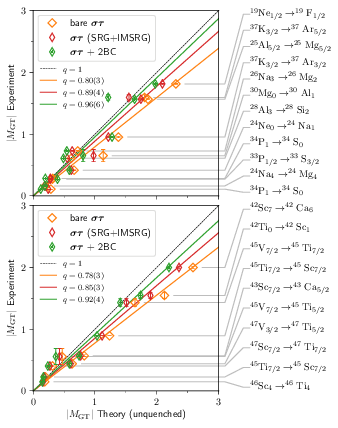}}
  \caption{Comparison of experimental ($y$-axis) and theoretical ($x$-axis) Gamow-Teller matrix elements for medium-mass nuclei. The theoretical results were obtained using (i) a bare Gamow-Teller one-body operator, (ii) Gamow-Teller one-body  operator consistently evolved with the Hamiltonian~\cite{Gysbers:2019uyb}, and (iii) a consistently-evolved Gamow-Teller operator that includes both one- and two-body currents. See Ref.~\cite{Gysbers:2019uyb} for details.}\label{fig:betamedium}
\vspace{-0.6cm}
\end{wrapfigure}
groups~\cite{Pastore:2017uwc,King:2020wmp,Gysbers:2019uyb}. The authors  in 
Refs.~\cite{Pastore:2017uwc,King:2020wmp}  calculated the Gamow-Teller matrix elements
in $A = 6-10$ nuclei accounting systematically for many-body effects in nuclear
interactions and coupling to the axial current both derived in $\chi$EFT.
The agreement of the calculations with the data is excellent and  the `$g_A$-problem'
can be resolved in light nuclei largely by correlation effects in the nuclear wave functions.
A summary of these calculations is reported in Fig.~\ref{fig:beta}. Similar results for these light
nuclei obtained using the No-core shell model are reported in Ref.~\cite{Gysbers:2019uyb}.

The $\chi$EFT approach in recent year is being implemented in studies of medium mass nuclei~\cite{Gysbers:2019uyb}. 
As a representative of this class of electroweak calculations we show the results of Ref.~\cite{Gysbers:2019uyb} on 
beta decay matrix elements represented in Fig.~\ref{fig:betamedium}. Here, the authors demonstrate that the quenching in the nuclear matrix elements
arises primarily from $\chi$EFT axial two-body currents and strong correlations in the nucleus. 
Nuclei from $A=3$ to $^{100}$Sn are analyzed with $\chi$EFT predictions in agreement with experimental data.

There has been exceptional progress in studying nuclear physics using $\chi$EFT. In the last two decades, this 
framework rooted in the symmetries and breaking pattern of QCD has allowed for the calculation of many low-energy nuclear processes, such as electromagnetic 
reactions and $\beta$ decays in both light and medium-mass nuclei, has reached a remarkable 
agreement with experiment, and has contributed to solving long-standing anomalies in nuclear theory. 
As chiral interactions and currents are being refined and pushed to higher orders, we have entered
the precision era  of this powerful framework.

\bibliographystyle{unsrtnat}
\bibliography{beta.bib}
\end{document}